\documentclass[twocolumn,english]{revtex4-2}
\usepackage[utf8]{inputenc}
\usepackage[T1]{fontenc}
\usepackage{hyperref}
\hypersetup{colorlinks=true, linkcolor=blue, filecolor=magenta, urlcolor=cyan,}
\usepackage{amsmath}
\usepackage{amsfonts}
\usepackage{amssymb}
\usepackage[version=4]{mhchem}
\usepackage{stmaryrd}
\usepackage{mathrsfs}
\usepackage{graphicx}
\usepackage[export]{adjustbox}

\begin{document}
\title{Frustrations in the ground state of a dilute Ising chain in a magnetic field }

\author{Y.D. Panov}

\affiliation{Institute of Natural Sciences, Ural Federal University, Yekaterinburg, Russia}

\begin{abstract}
The properties of the ground state of one of the simplest models of frustrated magnetic systems, a dilute Ising chain in a magnetic field, are considered for all values of the concentration of charged non-magnetic impurities. An analytical method is proposed for calculating the residual entropy of frustrated states, including states at the boundaries between the phases of the ground state, which is based on the Markov property of the system under consideration and allows direct generalization to other one-dimensional spin models with Ising-type interactions. The properties of local distributions and concentration dependences of the composition, correlation functions, magnetization and entropy of the phases of the ground state of the model are investigated. It is shown that the field-induced transition from the antiferromagnetic ground state to the frustrated one is accompanied by charge ordering and the absence of pseudo-transitions in the dilute Ising chain is proved.
\end{abstract}


\maketitle

\section{Introduction}
Absence or difficulty of long-range order formation is the basis of unusual behavior of low-dimensional spin and pseudospin systems. One-dimensional spin chains, including one-dimensional generalized Ising models, which have become popular in recent years [1], feature frustrated phases in the ground state. Frustrated states may be associated with different exotic properties of these systems such as magnetization plateau, quasiphases or pseudotransitions detected in decorated Ising systems [2-12].

Impurities may be the source of frustrations in magnetics, besides the lattice geometry. The dilute Ising chain is the simplest system model where the ground state is frustrated due to the presence of impurities. In the zero magnetic field, this model has an exact solution [13]. Its various properties are thoroughly investigated in [14-16], and the exact solution is analyzed in the most general form by Balagurov, Vaks and Zaitsev [17]. Taking into account the magnetic field, the standard transfer matrix method allows to study the thermodynamic properties of this model using the numerical solution of a nonlinear algebraic equation system. This method was used to address the entropy and Gruneisen magnetic parameter at finite temperatures [18]. However, the ground state properties, in particular concentration dependences of various physical values, may be only investigated within the standard method at the qualitative level from the numerical solution review at low temperatures.

An analytical calculation method is offered herein for various physical properties of the ground state of the dilute Ising chain in a longitudinal magnetic field at all possible model parameter values. The set of states at the ground state phase boundaries was defined according to the maximum residual entropy concept whose explicit expression was derived from the Markov property of the considered system [19]. The offered method allows obvious generalizations for one-dimensional pseudospin models with anisotropic interactions such as Ising, Potts, Blume-Capel and Blume-Emery-Griffiths. The ground state of the dilute Ising chain in the longitudinal magnetic field and its transformations induced by the magnetic field were investigated thoroughly and, in particular, it was shown that an unusual magnetoelectric effect occurs at certain parameters when a change in external magnetic field results in ordering of non-magnetic impurities.

The paper is organized as follows. In section 2, ground state phase diagrams are plotted and studied, correlation functions and properties of the local distributions of frustrated ground state phases are obtained. In section 3, the general equation is derived and concentration dependences are found for the residual entropy of the frustrated phases. The method to determine the set of states and residual entropy at the phase boundaries is described in Section 4, where comparison with the exact solution in zero field is also given. In section 5, magnetization for the phase boundary states is discussed. Summary is given in Section 6 .

\section{The ground state diagram}
The ground state of the dilute Ising chain in the zero magnetic field is described in [19] and qualitatively described in [18] taking into account the magnetic field. This Section offers a strict procedure for obtaining phase diagrams of the model ground state with fixed impurity concentration in the external magnetic field.

The model Hamiltonian may be written as [18]:
\begin{equation}
	\mathcal{H} = 
	- J \sum_{j=1}^{N} S_{z,j} S_{z,j+1} 
	+ V \sum_{j=1}^{N} P_{0,j} P_{0,j+1} 
	- h \sum_{j=1}^{N} S_{z,j} . 
	\label{eq:H1}
\end{equation}
Pseudospin operator $S=1$ is used herein, where the spin doublet and nonmagnetic impurity states correspond to projections $S_{z}= \pm 1$ and $S_{z}=0, J$ is the exchange interaction constant, $V>0$ is the effective inter-site interaction for impurities, $P_{0}=1-S_{z}^{2}$ is the projection operator on the impurity state. Fixed concentration of nonmagnetic impurities $n=\left\langle\sum_{j} P_{0, j}\right\rangle / N$ is assumed. Further we will assume an annealed system case. Interaction with $V=V_{0}+V_{1}-2 V_{01}$ is known to be equivalent to a more general interaction:
$$
\begin{aligned}
V_{0} \sum_{j} P_{0, j} P_{0, j+1} & +V_{1} \sum_{j} P_{s, j} P_{s, j+1} \\
& +V_{01} \sum_{j}\left(P_{0, j} P_{s, j+1}+P_{s, j} P_{0, j+1}\right),
\end{aligned}
$$
where $P_{s}=S_{z}^{2}$ is the projector on magnetic states.

For pre-defined $n$, the system energy may be expressed in the form of a sum over bonds. Write $N_{a, b}$ for the number of bonds with the left site in state $a$ and with the right site in state $b$, where $a, b=1,0,-1$, so $\sum_{a b} N_{a, b}=N$, and calculate concentrations $x_{\alpha}$ by expressions $x_{a, a}=N_{a, a} / N$, $x_{a, b}=\left(N_{a, b}+N_{b, a}\right) / N$, $a \neq b$. Then, index $\alpha$ corresponds to the unordered pair $(a, b)$, and $\sum_{\alpha} x_{\alpha}=1$. Values of $x_{\alpha}$ generally depend on temperature and all other model parameters and are expressed through the corresponding pair distribution functions [19]. In the ground state, the system energy per site, $\varepsilon=E / N$, is the linear function of variables $x_{\alpha}$ :
\begin{equation}
	\begin{aligned}
	\varepsilon= & -J\left(x_{1,1}+x_{-1,-1}-x_{1,-1}\right)+V x_{0,0} \\
	& -h\left(x_{1,1}-x_{-1,-1}+\frac{1}{2}\left(x_{0,1}-x_{0,-1}\right)\right) .
	\end{aligned}
\end{equation}
Search for minimum ground state energy is reduced to the solution of the canonical linear programming problem 
\begin{equation}
	\left\{\begin{array}{l}
	\varepsilon\left(x_{\alpha}\right) \rightarrow \min , \\
	x_{0,0}+\frac{1}{2}\left(x_{0,1}+x_{0,-1}\right)=n, \\
	x_{1,1}+x_{-1,-1}+x_{1,-1}+\frac{1}{2}\left(x_{0,1}+x_{0,-1}\right)=n_{s}, \\
	x_{\alpha} \geq 0,
	\end{array}\right.
\end{equation}
where $n_{s}=1-n$ is the spin sites concentration.

Solutions of problem (3) correspond to the vertices, edges or faces of the feasible solution polyhedron of $x_{\alpha}$. Solutions on vertices are listed in Table I. They define the ground state phase existence regions in the diagram shown in Figure 1. Hereinafter $m$ denotes the deviation from the half-filling for the concentration of impurities, $m=n-1 / 2$. The magnetization is defined by extression
\begin{equation}
	M=x_{1,1}-x_{-1,-1}+\frac{1}{2}\left(x_{0,1}-x_{0,-1}\right) .
\end{equation}

Solutions from 1 to 3 exist at all $n, 0 \leq n<1$. In the absence of impurities at $n=0$, only ferromagnetic (FM) ordering (solutions 1 and 2) and antiferromagnetic (AFM) ordering (solution 3) are implemented which are separated by the critical field $|h|=-2 J$ (the spin-flip field). Phase diagram for this case is shown in Figure 1, $a$. At $n \neq 0$, solutions 1 and 2 describe FM phases where macroscopic domains of ferromagnetically ordered spins directed along the field are separated by the nonmagnetic impurity domains. In this case, $x_{0,0} \neq 0$ and $x_{ \pm 1, \pm 1} \neq 0$, and $x_{0, \pm 1}=0$ in the thermodynamic limit. FM phases have the lowest energy at $J>V>0, h \neq 0$. Magnetization of FM phases is equal to the spin site concentration, $M=n_{s}$. AFM phase 3 occurs at $J<-V-|h|$ and consists of the alternating macroscopic domains of antiferromagnetically ordered spins and impurity domains, and has zero magnetization.

Solutions from 4 to 7 exist only for low-dilute spin chain, $0<n<1 / 2$, and their energies do not depend on $V$ (see Figure 1, $b$ ). Concentrations $x_{0,0}=0$ and $x_{0, \pm 1}=2 n$ show that dilute AFM or FM state is implemented where (A)FM spin clusters of different sizes, including single spins, are separated by single nonmagnetic impurities. As shown below, these solutions have nonzero residual entropy, therefore phases 4 and 5 may be called frustrated ferromagnetic phases (FR-FM), and phases 6 and 7 as frustrated antiferromagnetic phases (FR-AFM). Expressions for the residual entropy are also listed in Table I. At $n=1 / 2$, charge-ordered state occurs where spin and impurity sites alternate and the energy does not depend on the interaction constants $J$ and $V$. Magnetization in FR-FM phases is equal to the spin center concentration, $M=n_{s}$, and decreases with increasing $n$, and in FR-AFM phases $M=n$.

Solutions No. 8 and 9 exist only for the high-dilute spin chain, $1 / 2 \leq n<1$, at $-V-|h|<J<V$ (see Figure $1, c$ ). For these solutions $x_{ \pm 1, \pm 1}=0, x_{1,-1}=0, x_{0,0}=2 m$, and $x_{0, \pm 1}=2 n_{s}$, which corresponds to frustrated paramagnetic phases (FR-PM). In these phases, single spins directed along the filed are separated by the impurity clusters of different sizes, magnetization $M=n_{s}$ and energy does not depend on $J$.

For solutions 10 and 11, energy is always higher than the minimum energy at $h \neq 0$, but, as will be shown below, these solutions are included in the states at the phase boundary $h=0$.

\begin{figure*}
\includegraphics[width=0.8\textwidth]{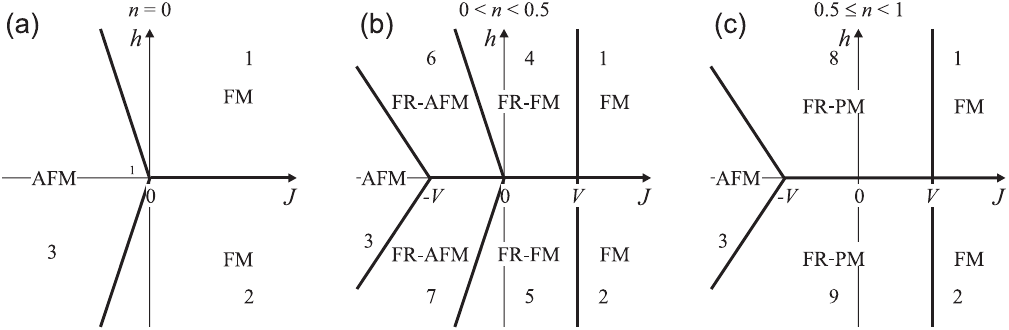}
\caption{Ground state phase diagrams of the dilute one-dimensional Ising model in the longitudinal magnetic field in $(h, J)$-plane for $(a)$ pure spin chain, $n=0 ;(b)$ low-dilute spin chain, $0<n<1 / 2 ;(c)$ high-dilute spin chain, $1 / 2<n<1$. The digits in the diagram correspond to the solutions in Table I.}
\end{figure*}

{
\renewcommand{\arraystretch}{1.3}
\renewcommand{\tabcolsep}{1em}
\begin{table*}[th]
\caption{The set of $\left\{x_{\alpha}\right\}$ problem solutions (3) in the vertices of polyhedron, feasible solutions and corresponding values of residual entropy $s_{0}$. 
\label{tab:phases}
}
	\begin{tabular}{ c c c c c c c c c }
\hline
		 Solution &  & $\varepsilon$ 
		& $x_{0,0}$ & $x_{1,1}$ & $x_{-1,-1}$ & $x_{1,-1}$ & $x_{0,1}$ & $x_{0,-1}$ \\ 
\hline
		1 & & $-\left(J+h\right) n_s+V n$ & $n$ & $n_s$ & $0$ & $0$ & $0$ & $0$ \\
		2 & & $-\left(J-h\right) n_s+V n$ & $n$ & $0$ & $n_s$ & $0$ & $0$ & $0$ \\
		3 & & $J n_s+V n$ & $n$ & $0$ & $0$ & $n_s$ & $0$ & $0$ \\[1em]
 & \multicolumn{8}{c}{$s_0 = 0$ }\\[1em]
\hline
		4 & $m<0$ & $2Jm-h n_s$ & $0$ & $-2m$ & $0$ & $0$ & $2n$ & $0$ \\
		5 & $m<0$ & $2Jm+h n_s$ & $0$ & $0$ & $-2m$ & $0$ & $0$ & $2n$ \\[1em]
 & \multicolumn{8}{c}{$s_0 = - 2|m| \ln\left(2|m|\right) - \left(\frac{1}{2}-|m|\right) \ln\left(\frac{1}{2}-|m|\right) 
	+ \left(\frac{1}{2}+|m|\right) \ln\left(\frac{1}{2}+|m|\right)$ }\\[1em]
\hline
		6 & $m<0$ & $-2Jm-h n$ & $0$ & $0$ & $0$ & $-2m$ & $2n$ & $0$ \\
		7 & $m<0$ & $-2Jm+h n$ & $0$ & $0$ & $0$ & $-2m$ & $0$ & $2n$ \\[1em]
 & \multicolumn{8}{c}{$s_0 = - |m|\ln|m| - \left(\frac{1}{2}-|m|\right)\ln\left(\frac{1}{2}-|m|\right) - \frac{1}{2}\ln2$ } \\[1em]
\hline
		8 & $m\geq0$ & $2Vm-h n_s$ & $2m$ & $0$ & $0$ & $0$ & $2n_s$ & $0$ \\
		9 & $m\geq0$ & $2Vm+h n_s$ & $2m$ & $0$ & $0$ & $0$ & $0$ & $2n_s$ \\[1em]
 & \multicolumn{8}{c}{$s_0 = - 2|m| \ln\left(2|m|\right) - \left(\frac{1}{2}-|m|\right) \ln\left(\frac{1}{2}-|m|\right) 
	+ \left(\frac{1}{2}+|m|\right) \ln\left(\frac{1}{2}+|m|\right)$ }\\[1em]
\hline
		10 & $m\leq0$ & $2\left(J+h\right) m+h n$ & $0$ & $-2m$ & $0$ & $0$ & $0$ & $2n$ \\
		11 & $m\leq0$ & $2\left(J-h\right) m-h n$ & $0$ & $0$ & $-2m$ & $0$ & $2n$ & $0$ \\ \hline
	\end{tabular}
\end{table*}
}

Transition from the AFM phase to FR-AFM or FR-PM phase at the specified impurity concentration may be caused by the magnetic field variation. Filed $|h|=-J-V$ (where $J<-V<0$ ) that defines the boundary between AFM and frustrated phases may be called frustration field. When the magnetic field exceeds the frustration field, an unusual effect occurs: charge ordering occurs in the system
caused by the change in the magnetic field. Nonzero $x_{0, \pm 1}$ in FR-AFM and FR-PM phases (see Table I) are the markers of charge ordering, while in AFM phase $x_{0, \pm 1}=0$. The charge order reaches its maximum at half filling, $m=0$, and in this case the ground state variation will be manifested more clearly: dilute AFM state at $|h|<-J-V$ consisting of macroscopic AFM domains and impurity domains and having nonzero magnetization is followed by charge-ordered state at $|h|>-J-V$, where spin and impurity centers alternate and magnetization is equal to $M=1 / 2$.

Using the data in Table I, the correlation functions may be defined and local state distribution characteristics in the spin chain can be calculated [19] to supplement the description of the system phase states.

Pair distribution functions $\left\langle P_{a, k} P_{b, k+l}\right\rangle$, where $P_{a, k}$ is the projector on state $a=1,0,-1$ at site $k$, can be found [19] using the conditional probability matrix $\mathcal{P}_{a b}=P(a|b)$:
\begin{equation}
	\left\langle P_{a,k} P_{b,k+l} \right\rangle = \mathcal{P}_{ab}^{l} P(b) . 
	\label{eq:ChapKol}
\end{equation}
Conditional probability $P(a|b)$ that the $i$-th site in state $a$, provided that the $(i+1)$-th site is in state $b$, is defined by the type of bond. If $a=b$, then $x_{a, a}=P(a a)=P(a) P(a|a)$. Probabilities $P(a)$ are equal
to the site concentrations in the corresponding states, $P(a)=x_{a, a}+\tfrac{1}{2} \sum_{b \neq a} x_{a, b}$. Considering the equality of the two directions along the chain, we obtain that for $a \neq b$ equalities $N_{a, b}=N_{b, a}=\tfrac{1}{2} x_{a, b} N$ and $P(a b)=P(b a)$ shall be satisfied, whence it follows that $x_{a, b}=2 P(a b)=2 P(a|b) P(b)$. Next, correlation functions 
$K_{a b}(I)=\left\langle P_{a, k} P_{b, k+l}\right\rangle-\left\langle P_{a}\right\rangle\left\langle P_{b}\right\rangle$, and, taking into account $S_{z, k}=P_{1, k}-P_{-1, k}$, spin correlation function 
$C(l)=\left\langle S_{z, k} S_{z, k+l}\right\rangle-\left\langle S_{z}\right\rangle^{2}$ can be calculated.

For FR-FM phase at $h>0$ and $m<0$, state $a=-1$ is not available, 
$P(1)=\tfrac{1}{2}+|m|$, $P(0)=\tfrac{1}{2}-|m|$. The conditional probability matrix at states $a=1,0$ is written as
\begin{equation}
	\mathcal{P} = 
	\begin{pmatrix}
	\frac{4|m|}{1+2|m|} & 1 \\
	\frac{1-2|m|}{1+2|m|} & 0 
	\end{pmatrix} . 
	\label{eq:P-frfm}
\end{equation}
Using (5), find the pair distribution functions for impurities and for spins
\begin{eqnarray}
	\left\langle P_{0,k} P_{0,k+l} \right\rangle 
	&=& \left(\frac{1}{2} - |m|\right)^{2} \nonumber\\
	&&+ (-1)^l \left(\frac{1}{4}-m^2\right) \left(\frac{1-2|m|}{1+2|m|}\right)^{l} , \\ 
	\left\langle S_{z,k} S_{z,k+l} \right\rangle 
	&=& \left(\frac{1}{2} + |m|\right)^{2} \nonumber\\
	&&+ (-1)^l \left(\frac{1}{4}-m^2\right) \left(\frac{1-2|m|}{1+2|m|}\right)^{l} . 
\end{eqnarray}
The first terms in these expressions are equal to $\left\langle P_{0}\right\rangle^{2}$ and $\left\langle S_{z}\right\rangle^{2}$, respectively, therefore
$$
	K_{00}(l) = C(l) = (-1)^l \left(\frac{1}{4}-m^2\right) e^{-\frac{l}{\xi}} , \quad 
$$
\begin{equation}
	\xi = \left[\ln \left(\frac{1+2|m|}{1-2|m|}\right) \right]^{-1} . 
	\label{eq:xi-FRFM} 
\end{equation}
Correlation length $\xi$ becomes infinite at $m=0$, i.e. when the chain is half filled with impurities.

For $\mathrm{FR}-\mathrm{AFM}$ phase at $h>0$ and $m<0$, we have $P(1)=1 / 2, P(0)=1 / 2-|m|, P(-1)=|m|$, and the conditional probability matrix at states $a=1,0,-1$ is written as
\begin{equation}
	\mathcal{P} = 
	\begin{pmatrix}
	0 & 1 & 1 \\
	1-2|m| & 0 & 0 \\
	|m| & 0 & 0 
	\end{pmatrix} . 
\end{equation}
The corresponding correlation functions are characterized by the infinite correlation length
$$
	K_{00}(l) = (-1)^{l}\left(\frac{1}{2} - |m|\right)^{2} , \quad 
$$
\begin{equation}
	C(l) = (-1)^{l}\left(\frac{1}{2} + |m|\right)^{2} . 
\end{equation}

It should ne noted that in the zero field [19] at $|J|<V$, the impurity correlation functions are similar for $J>0$ and $J<0$, and the correlation length at $T=0$ for impurities is described by expression (9). The spin correlation length $\xi_{s 0}$ is also infinite, but has an other value
$$
	C_{0}(l) = (\pm1)^{l} \left(\frac{1}{2} + |m|\right) e^{-\frac{l}{\xi_{s0}}} , \quad 
$$
\begin{equation}
	\xi_{s0} = \left[\ln \left(\frac{1+2|m|}{4|m|}\right) \right]^{-1} , 
\end{equation}
where +1 and -1 correspond to $J>0$ and $J<0$.

For FR-FM phase at $h>0$ and $m>0$, state $a=-1$ is not available, $P(1)=1 / 2+|m|, P(0)=1 / 2-|m|$. The conditional probability matrix at states $a=1,0$ is written as
\begin{equation}
	\mathcal{P} = 
	\begin{pmatrix}
	 0 & \frac{1-2|m|}{1+2|m|} \\ 
	 1 & \frac{4|m|}{1+2|m|} 
	\end{pmatrix} , 
\end{equation}
and the expressions for the correlation functions of $\mathrm{FR}-\mathrm{PM}$ and FR-FM phases are the same. At FR-PM phase boundary at $h=0$ and $m>0$, the impurity correlation function remains the same, but the spin correlation function is equal to zero [19].

In addition, consider the characteristics of the local state distribution over the chain sites. Write $\sigma$ for the ordered set of $k$ adjacent sites in the specified states: $\sigma \equiv a_{1}, a_{2} \ldots a_{k}$. Let $(\sigma)$ be the sequence of some number of repeating blocks $\sigma$, its lengths is written as $l_{(\sigma)}$. Probabilities $p\left(l_{(\sigma)}\right)$ of this value $l_{(\sigma)}$ obey [19] the geometrical distribution
\begin{equation}
	p(l_{(\sigma)}) = \left(1-q_{(\sigma)}\right) q_{(\sigma)}^{l-1} , 
\end{equation}
where $l$ is the number of blocks in the sequence $(\sigma)$, and $q_{(\sigma)}$ has the meaning of the probability of cycle from states $\sigma$:
\begin{equation}
	q_{(\sigma)} = P(a_{1}|a_{2}) \dots P(a_{k-1}|a_{k}) P(a_{k}|a_{1}) . 
	\label{eq:q-sigma}
\end{equation}
The length of sequence $(\sigma)$ and its dispersion are written as
\begin{equation}
	\bar{l}_{(\sigma)} = \frac{1}{1-q_{(\sigma)}} , \qquad 
	D(l_{(\sigma)}) = \frac{q_{(\sigma)}}{\left(1-q_{(\sigma)}\right)^2} . 
\end{equation}

For example, find in FR-FM phase at $h>0$ the mean length of spin sequence which is written as $(\sigma)=(1)$ in this case. Using equations (6) and (15), we get
\begin{equation}
	q_{(1)} = \frac{4|m|}{1+2|m|} , \qquad
	\bar{l}_{(1)} = \frac{1+2|m|}{1-2|m|} . 
\end{equation}
The mean spin sequence length is maximum at $m=-1 / 2$ and reaches its minimum $\vec{l}_{(1)}=1$ at half filling, $m=0$. For the impurity sequence $l_{(0)}=1$, which corresponds to the isolated impurities. 
For the charge-ordered sequence $(\sigma)=(01)$, we get
\begin{equation}
	\bar{l}_{(01)} = \frac{1+2|m|}{4|m|} . 
	\label{eq:l01-frfm}
\end{equation}
As can be seen $\bar{l}_{(01)}$ becomes infinite at the half filling.

In FR-PM phase at $h>0$ and $m>0$, states 0 and 1 change places: in this case, single spins directed along he field are separated by the impurity sequences with the mean length
$$
\bar{l}_{(0)}=\frac{1+2|m|}{1-2|m|} . 
$$
Expression (18) for $\bar{l}_{(01)}$ is preserved.

In FR-AFM phase, any sequence $(\sigma)=(a)$, where $a=1,\,0,\,-1$, has the minimum length $l_{(a)}=1$, and for the antiferromagnetic sequence $(\sigma)=(-11)$, the mean length
\begin{equation}
	\bar{l}_{(-11)} = \frac{1}{1-2|m|} 
\end{equation}
becomes infinite when no impurities are available. However, when the spin states are combined in a single state $s$ using projector $P_{s}=P_{1}+P_{-1}$, then the picture will change. For the spin sequence $(\sigma)=(s)$ and for charge-ordered sequence $(\sigma)=(0 \mathrm{~s})$, we get
\begin{equation}
	\bar{l}_{(s)} = \frac{1+2|m|}{1-2|m|} , \qquad 
	\bar{l}_{(0s)} = \frac{1+2|m|}{4|m|} . 
\end{equation}
From this point of view, FR-FM and FR-AFM phases are equivalent. Note that these properties of local distributions are also preserved at the zero field [19].

\section{General expression for the residual entropy of the dilute Ising chain in magnetic field}

Using the Markov property of the dilute Ising chain [19], 
write the state probability $a_{1}, a_{2} \ldots a_{N}$ of the closed chain of $N$-sites $(N \gg 1)$ :
\begin{eqnarray}
	P_{\mathcal{O}}(a_1 a_2 \ldots a_N) 
	&=& P(a_1|a_2)P(a_2|a_3) \ldots P(a_N|a_1) = \nonumber \\ 
	&=& \prod_{ab} P(a|b)^{N_{ab}} = p_0^N ,
	\label{eq:P0}
\end{eqnarray}
where
\begin{multline}
	p_0 = \left(\frac{x_{0,0}}{P(0)}\right)^{x_{0,0}}
	\left(\frac{x_{1,1}}{P(1)}\right)^{x_{1,1}} 
	\left(\frac{x_{-1,-1}}{P(-1)}\right)^{x_{-1,-1}} 
	\\ \times
	\bigg(\frac{x_{1,-1}^2}{4P(1)P(-1)}\bigg)^{\!\!\!\frac{x_{1,-1}}{2}}
	\!\!\bigg(\frac{x_{0,1}^2}{4P(0)P(1)}\bigg)^{\!\!\!\frac{x_{0,1}}{2}}
	\!\!\bigg(\frac{x_{0,-1}^2}{4P(0)P(-1)}\bigg)^{\!\!\!\frac{x_{0,-1}}{2}}. 
	\label{eq:p00}
\end{multline}

The ground state energy (2) is defined by values $x_{\alpha}$. Assuming that a microcanonical distribution holds true for the ground state, find the statistical weight $\Gamma$ of the ground state and residual entropy $s_{0}$ :
\begin{equation}
	\Gamma = P_{\mathcal{O}}^{-1} , \quad
	s_0 = \frac{\ln\Gamma}{N} = - \ln p_0 .
\end{equation}
Considering (22), we get
\begin{equation}
	s_0 = - \sum_{\alpha} x_{\alpha} \ln x_{\alpha} + P_2 \ln2 + \sum_a P(a) \ln P(a) , 
	\label{eq:s0-common}
\end{equation}
where the total concentration of pairs from different states 
$P_{2}=x_{1,-1}+x_{0,1}+x_{0,-1}$ is introduced.

Equation (24) allows to find the concentration dependence of the residual entropy at the given ground state set $\left\{x_{\alpha}\right\}$. To solve this problem within the standard approach, we need to find the maximum transfer matrix eigenvalue, determine the parametrical dependence of the entropy on concentration using the chemical potential as a parameter, and find the limit at the zero temperature. For the dilute Ising chain in magnetic filed, this can be made only numerically [18], while equation (24) gives the exact analytical result.

Table I shows the expressions for the residual entropy of various ground state phases. FM and AFM solutions have the zero entropy. Solutions from 4 to 9 have nonzero residual entropy for all impurity concentrations, except $n=0,0.5$ and 1 . Note that for FR-FM and FR-PM solutions, the entropy has the same dependence on $|m|$ demonstrating a kind of symmetry of impurity and spin states in FM phases and, at the specified concentration, this value is higher than the entropy of FR-AFM states.

Concentration dependences of the residual entropy for solutions from 1 to 9 are shown in Figure 2,a. The obtained dependences agree with the entropy behavior at low temperatures that was obtained by numerical solution of the nonlinear algebraic equation system with a grand canonical ensemble [18]. The method described herein allows to study the residual entropy behavior analytically. For FR-FM and FR-PM phases, the entropy has maxima
$s_{0, \max }=-\frac{1}{2} \ln \frac{\sqrt{5}-1}{\sqrt{5}+1} \approx 0.481$ by 
$ m= \pm \frac{1}{2 \sqrt{5}} \approx \pm 0.224$,
and for FR-AFM phase - the maximum is
$s_{0,max} = \frac{1}{2} \ln 2 \approx 0.347$ 
achieved when $m = -\frac{1}{4}$.

\section{Residual entropy of states at the phase boundary}

Pseudotransitions are an outstanding feature of decorated Ising 1D models and are associated with the presence of frustrated phases in the ground state of these systems [2-12]. A stepwise change in the one-dimensional system state occurs during the pseudotransition at a finite temperature, as a result some thermodynamic functions demonstrate very sharp features, though remain continuous. Entropy and magnetization in the magnetic field are characterized by the stepwise dependence on temperature, and the heat capacity, susceptibility and correlation length have clearly pronounced maxima. Despite the common phase transition, the system state at the temperatures above the pseudotransition point is the frustrated phase which is more favorable due to the entropy contribution to the free energy. The pseudotransition temperature is the function of system parameters, including the magnetic field. This suggests the 
pseudotransition and related thermal effect can be controlled using the magnetic field.

\begin{figure*}
\includegraphics[width=0.8\textwidth]{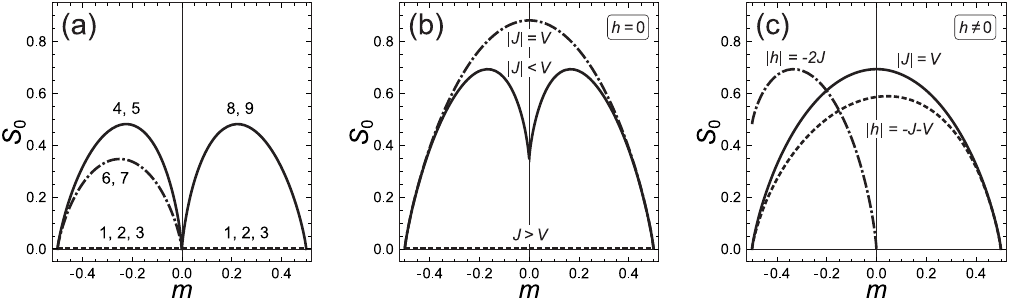}
\caption{Concentration dependences of the residual entropy of the dilute Ising chain for $(a)$ phase states listed in Table $I,(b, c)$ states at the phase boundary at $h=0$ and at $h \neq 0$ listed in Table II.}
\end{figure*}

To predict the pseudotransition, knowledge of exact residual entropy values for all system parameters and, in particular, at the phase boundaries in the ground state phase diagram is critical. According to the Rojas criterion [10,11], pseudotransition occurs near the frustrated phase boundary, if the entropy at the boundary itself is equal to the frustrated phase entropy. Such situation is rather uncommon, therefore the mere existence of frustrated phase does not mean that a pseudotransition exists.

The adjacent phase energies in the phase diagram become equal at the phase boundary, therefore the state at the boundary shall be a mixture of the adjacent phases, unless such mixture results in the system energy increase. Define coefficients $c_{n}$ as the variables in linear combinations $x_{\alpha}=\sum_{n} c_{n} x_{\alpha}^{(n)}$, where $x_{\alpha}$ are unknown concentrations for the state at the boundary, and $x_{\alpha}^{(n)}$ are the found concentrations for the adjacent phases. Coefficients $c_{n}$ will be found according to the maximum entropy concept. Using equation (24) for $s_{0}$, a nonlinear optimization problem is derived
\begin{equation}
	\left\{
	\begin{array}{l}
		s_0(c_{n}) \rightarrow \max , \\ 
		\sum_{n} c_{n} = 1 , \\
		c_{n} \geq 0 . \\ 
	\end{array}
	\right.
	\label{eq:S0max}
\end{equation}

{\renewcommand{\arraystretch}{1.1}
\setlength{\tabcolsep}{1em}
\begin{table*}[th]
\caption{
Set of $\left\{x_{\alpha}\right\}$ and residual entropy $s_{0}$ at ground state phase boundaries. Phase numbers included in the linear combination are listed in the first column, the necessary condition for solution existence is listed in the second column.
\label{tab:border}
}

\begin{tabular}{ c c c c c c c c }\hline
	Solution 
	&  
	& $x_{0,0}$ & $x_{1,1}$ & $x_{-1,-1}$ & $x_{1,-1}$ & $x_{0,1}$ & $x_{0,-1}$ \\ 
\hline
	& $h=0$ & \multicolumn{6}{c}{ } \\
	1, 2 
	& $J>V$ 
	& $n$ 
	& $\tfrac{1}{2} n_s$ 
	& $\tfrac{1}{2} n_s$ 
	& $0$ & $0$ & $0$ \\[0.5em] 
	  \multicolumn{8}{c}{$s_0 = 0$ }\\[0.5em]
\hline
	& $h=0$ & \multicolumn{6}{c}{ } \\
	$4, 5, 10, 11$ 
	&  $0<J<V$
	& $0$ & $-m$ & $-m$ & $0$ & $n$ & $n$ \\
	6, 7 
	&  $-V<J<0$
	& $0$ & $0$ & $0$ & $-2m$ & $n$ & $n$ \\
	8, 9 
	&  $|J|<V$
	& $2m$ & $0$ & $0$ & $0$ & $n_s$ & $n_s$ \\[0.75em] 
	  \multicolumn{8}{c}{$s_0 =
	- 2|m| \ln\left(2|m|\right)
	- \left(\frac{1}{2}-|m|\right) \ln\left(\frac{1}{2}-|m|\right) 
	+ \left(\frac{1}{2}+|m|\right) \ln\left(\frac{1}{2}+|m|\right) 
	+ \left(\frac{1}{2}-|m|\right) \ln2$ }\\[0.75em]
\hline
	& $h=0$ & \multicolumn{6}{c}{ } \\
	$1,2,4,5,10,11$ 
	& $J=V$
	& $x^{\ast}+m$ 
	& $\tfrac{x^{\ast}-m}{2}$ 
	& $\tfrac{x^{\ast}-m}{2}$ 
	& $0$ & $\tfrac{1}{2}-x^{\ast}$ & $\tfrac{1}{2}-x^{\ast}$ \\
	 $1, 2, 8, 9$ 
	& $J=V$
	& $x^{\ast}+m$ 
	& $\tfrac{x^{\ast}-m}{2}$ 
	& $\tfrac{x^{\ast}-m}{2}$ 
	& $0$ & $\tfrac{1}{2}-x^{\ast}$ & $\tfrac{1}{2}-x^{\ast}$ \\
	3, 6, 7 
	& $J=-V$
	& $x^{\ast}+m$ & $0$ & $0$ 
	& $x^{\ast}-m$ & $\tfrac{1}{2}-x^{\ast}$ & $\tfrac{1}{2}-x^{\ast}$ \\
	 3, 8, 9 
	& $J=-V$
	& $x^{\ast}+m$ & $0$ & $0$ 
	& $x^{\ast}-m$ & $\tfrac{1}{2}-x^{\ast}$ & $\tfrac{1}{2}-x^{\ast}$ \\[0.5em] 
	  \multicolumn{8}{c}{$
	s_0 = \left(\frac{1}{2}+m\right) \ln \frac{1+2m}{2x^{\ast}+2m}
	+\left(\frac{1}{2}-m\right) \ln \frac{1-2m}{2x^{\ast}-2m}
	$ }\\[0.5em]
\hline
	& $J=V$ & \multicolumn{6}{c}{ } \\
	1, 4 
	& $h>0$ 
	& $n^2$ & $n_s^2$ & $0$ & $0$ & $2nn_s$ & $0$ \\ 
	2, 5 
	& $h<0$ 
	& $n^2$ & $0$ & $n_s^2$ & $0$ & $0$ & $2nn_s$ \\ 
	1, 8 
	& $h>0$ 
	& $n^2$ & $n_s^2$ & $0$ & $0$ & $2nn_s$ & $0$ \\ 
	2, 9 
	& $h<0$ 
	& $n^2$ & $0$ & $n_s^2$ & $0$ & $0$ & $2nn_s$ \\[0.5em]
	  \multicolumn{8}{c}{ $s_0 = - \left(\frac{1}{2} - m\right) \ln \left(\frac{1}{2} - m\right) 
	- \left(\frac{1}{2} + m\right) \ln \left(\frac{1}{2} + m\right)$ }\\[0.5em]
\hline
	& $J=-V-|h|$ & \multicolumn{6}{c}{ } \\ 
	3, 6
	& $h>0$ 
	& $x_0$ & $0$ & $0$ & $x_1$ & $x_2$ & $0$ \\ 
	3, 7
	& $h<0$ 
	& $x_0$ & $0$ & $0$ & $x_1$ & $0$ & $x_2$ \\ 
	3, 8
	& $h>0$ 
	& $x_0$ & $0$ & $0$ & $x_1$ & $x_2$ & $0$ \\ 
	3, 9
	& $h<0$ 
	& $x_0$ & $0$ & $0$ & $x_1$ & $0$ & $x_2$ \\[0.5em] 
	  \multicolumn{8}{c}{$s_0 = -\left(\frac{1}{2}+m\right) \ln \left(1-\mu\alpha\right) 
	+\frac{1}{2}\left(\frac{1}{2}-m\right) \ln \frac{1+\alpha}{1-\alpha}$ }\\[0.5em]
\hline
	& $J=-|h|/2$ & \multicolumn{6}{c}{ } \\ 
	4, 6 
	& $h>0$ 
	& $0$ & $-2m-x^{\ast\ast}$ & $0$ & $x^{\ast\ast}$ & $2n$ & $0$ \\ 
	5, 7 
	& $h<0$ 
	& $0$ & $0$ & $-2m-x^{\ast\ast}$ & $x^{\ast\ast}$ & $0$ & $2n$ \\[0.5em] 
	  \multicolumn{8}{c}{$s_0 = - \left(\frac{1}{2}+m\right) \ln \left(1+2m\right) 
	+ \frac{1}{2} \ln \left(1-2m-x^{\ast\ast}\right)
	+ m \ln x^{\ast\ast}$ }\\[0.5em]
\hline
	\end{tabular}
\end{table*}
}

Results for the ground state phase boundary at $h=0$ are listed in Table II. Note that to obtain the states at $m<0$, $0<J \leq V$, solutions 10 and 11 from Table I shall be considered. Compare these results with the exact solution for the zero field [19]. Define the transfer matrix for Hamiltonian 
$\tilde{\mathcal{H}}=\mathcal{H}-\mu \Sigma_{j} P_{0, j}$, $\mu$ is the chemical potential at $h=0$ as follows
\begin{equation}
	\mathcal{T} =
	\left(
	\begin{array}{ccc}
		e^{K}     &  e^{\xi/2}   &  e^{-K}     \\
		e^{\xi/2} &  e^{-W+\xi}  &  e^{\xi/2}  \\
		e^{-K}    &  e^{\xi/2}   &  e^{K} 
	\end{array}
	\right) ,
\end{equation}
where $K=\beta J$, $W=\beta V$, $\xi=\beta \mu$, $\beta=1 / \theta$ and $\theta=k_{\mathrm{B}} T$. Transformation
\begin{equation}
	U = \frac{1}{\sqrt{2}} 
	\left( 
	\begin{array}{ccc} 
	1   &  \eta_2   &  \eta_3   \\ 
	0   &  A\eta_2  &  B\eta_3  \\ 
	-1  &  \eta_2   &  \eta_3   
	\end{array} 
	\right) , 
\end{equation}
where $A B=-2$ and $\eta_{2,3}$ are normalization factors, diagonalizes $\mathcal{T}$:
\begin{equation}
	\tilde{\mathcal{T}} = U^{+} \mathcal{T} U 
	= \left( 
	\begin{array}{ccc}
	\lambda_1   &  0   &  0   \\ 
	0   &   \lambda_2  &  0   \\ 
	0   &   0   &  \lambda_3   
	\end{array} 
	\right) , 
\end{equation}
\begin{equation}
	\lambda_1 = 2 \sinh K , 
\end{equation}
\begin{multline}
		\lambda_{2,3} 
		= \cosh K + \frac{e^{-W+\xi}}{2} \\
		\mp \left[ 2 e^{\xi} + \left(  \cosh K - \frac{e^{-W+\xi}}{2}  \right)^{\!2} \right]^{1/2} . 
\end{multline}

The maximum eigenvalue $\lambda_{3}$ allows to write the grand potential $\Omega$ and impurity concentration $n$ as a function of chemical potential
\begin{equation}
	\Omega = N \omega
	= - N \theta \ln \lambda_3 , \quad 
	n = \frac{1}{\lambda_3}\frac{\partial\lambda_3}{\partial\xi} . 
\end{equation}
The latter expression results in quadratic equation for activity $e^{\xi}$ having roots
\begin{equation}
	\left(e^{\xi}\right)_{\pm} 
	= \frac{ 8 e^{2W} }{ 1 - 4 m^2 } \left( g \pm m \right)^2 , 
	\label{eq:roots}
\end{equation}
where
\begin{equation}
	g = \left[ m^2 +\left( \frac{1}{4} - m^2 \right) e^{-W} \cosh K \right]^{1/2} . 
	\label{eq:g}
\end{equation}
For an ideal system with $K=0$ and $W=0$ the following expression can be derived directly
$$
e^{\xi}=\frac{1+2 m}{\frac{1}{2}-m} ,
$$
which defines the root selection
\begin{equation}
	e^{\xi} = \left(e^{\xi}\right)_{+}
	= 2 e^{W} \frac{ g + m  }{  g - m  } \, \cosh K .
\end{equation}
This allows to exclude the chemical potential in the eigenvalue expressions
\begin{equation}
	\lambda_{2,3} = \frac{ 2g \mp 1 }{ g - m } \, \cosh K ,
\end{equation}
and to express the matrix elements $U$:
$$
\eta_{2}=\sqrt{\frac{1}{2}+m}, \quad \eta_{3}=\sqrt{\frac{1}{2}-m} , 
$$
$$
A \eta_{2}=-\sqrt{1-2 m}, \quad B \eta_{3}=\sqrt{1+2 m} .
$$

Using the found values, concentration dependences of all thermodynamic model parameters in the zero field can be found $[19]$, including the pair distribution functions
\begin{equation}
	\left\langle P_{a,k} \, P_{b,k+l} \right\rangle
	= \lim_{N\to\infty} 
	\frac{
	\mathop{\mathrm{Tr}} \left(  P_a \mathcal{T}^l P_b \mathcal{T}^{N-l}  \right)
	}{ \mathop{\mathrm{Tr}} \left( \mathcal{T}^{N} \right) } , 
	\label{eq:pdf-def}
\end{equation}
where $P_{a, k}$ is the projection operator on the site $k$ per one of the basis states $a= \pm 1,0$ corresponding to $S_{z}= \pm 1,0$.

In this case, pair distribution functions are required for the nearest neighbors, because $\left\langle P_{a, k} P_{b, k+1}\right\rangle=P(a b)$. These
functions are written as:
\begin{equation}
	\langle P_{0,k} P_{0,k+1} \rangle
	= \frac{ \left( 1 + 2m \right) \left(g + m\right) }{ 2g + 1 } , 
\end{equation}
\begin{equation}
	\langle P_{\pm1,k} P_{\pm1,k+1} \rangle 
	= \frac{ \left(1-2m\right) \left(g - m\right) e^K }{ 4 \left(2g + 1\right) \cosh K } , 
\end{equation}
\begin{equation}
	\langle P_{\pm1,k} P_{\mp1,k+1} \rangle 
	= \frac{ \left(1-2m\right) \left(g - m\right) e^{-K} }{ 4 \left(2g + 1\right) \cosh K } , 
\end{equation}
\begin{equation}
	\langle P_{0,k} P_{\pm1,k+1} \rangle = \langle P_{\pm1,k} P_{0,k+1} \rangle
	= \frac{ 1 - 4m^2 }{ 4 \left(2g + 1\right) } . 
\end{equation}
In the limit $T \rightarrow 0$, for $g$ included in these equations the following is derived from (33):
\begin{equation}
	g  \;\; \underset{T\to0}{\longrightarrow} \;\;
	\left\{
		\begin{array}{ll}
			\displaystyle
			\propto e^{ \frac{J-V}{2\theta} } , & J > V ; \\[0.5em]
			\displaystyle
			|m| , & |J| < V ; \\[0.5em]
			\displaystyle
			\sqrt{\frac{1}{2}\left( \frac{1}{4} + m^2 \right)} , & |J| = V . 
		\end{array}
	\right.
\end{equation}

Considering these expressions in the limit at $T \rightarrow 0$, we obtain all values of $x_{\alpha}$ listed in Table II for $h=0$. This confirms the correctness of equation (24) for the residual entropy and method (25) to find the state entropy at the phase boundary.

Solutions at the ground state phase boundaries are listed in Table II. The following notations is introduced:
$$
\begin{gathered}
x_{0}=\left(\frac{1}{2}+m\right)(1-\mu \alpha), \quad x_{1}=\left(\frac{1}{2}-m\right)(1-\alpha), \\
x_{2}=(1-2 m) \alpha, \quad
x^{*}=\sqrt{\frac{1}{2}+2 m^{2}}-\frac{1}{2}, \\
x^{* *}=\frac{1}{5}\left(\frac{1}{2}-9 m-\sqrt{\frac{1}{4}-9 m+m^{2}}\right) .
\end{gathered}
$$
Here, $\alpha$ is here defined from equation
\begin{equation}
	\left(1 - \mu \alpha\right)\sqrt{1 - \alpha^2} = 2 \mu \alpha^2 , \quad
	\mu = \frac{1-2m}{1+2m} . 
	\label{eq:alpha-eq}
\end{equation}
If $m>0$, then $0 \leq \alpha \leq 1$, and if $m<0$, then $0 \leq \alpha \leq 1 / \mu$.
Solutions in Table II are divided into groups having similar concentration dependences of entropy. The found concentration dependences of entropy at $h=0$ coincide with those obtained before [19] from the exact solution for the dilute Ising chain in the zero field in the limit at $T \rightarrow 0$, which again verifies correctness of equation (24). The state entropy at $h=0$ and $|J|<V$ has two maxima:
$s_{0, \max }=\ln 2 \approx 0.693$ by $m= \pm 1 / 6$ and a local minimum,
$s_{0, \text { max }}=\frac{1}{2} \ln 2 \approx 0.347$ by $m=0$.
At $h=0$ and $J=V$, the entropy has its maximum
$s_{0, \max }=\ln (1+\sqrt{2}) \approx 0.881$ by  $m=0$.

At the boundary between FM and frustrated phases, $J=V$, $h \neq 0$, the entropy is symmetric with respect to $m=0$ and has its maximum
$s_{0, \max }=\ln 2 \approx 0.693$ by $m=0$.

The state entropy at the boundary between AFM and frustrated phases, 
$|h|=-J-V$ at $J<-V<0$, is not symmetric with respect to $m=0$ and reaches its maximum 
$s_{0, \max } \approx 0.589$ at $m=0.043$. 
At the boundary corresponding to the spin flip field, 
$m<0$, $J=-|h| / 2$, $h \neq 0$, the entropy has its maximum 
$s_{0, \max }=\ln 2$ by $m=-\frac{1}{3}$.

The view of the concentration dependences of the residual entropy from Table II is shown in Figure 2, $b$ and $c$.

In all considered cases, the state entropy at the ground state phase boundaries is higher than the entropy of the adjacent phases. Using the Rojas criterion $[10,11]$, it can be concluded that the pseudotransition does not occur in the one-dimensional dilute Ising model.

\section{Magnetization of states at phase boundaries}
Magnetization of states at phase boundaries can be calculated by equation (4) using the solutions in Table II. The magnetization at the boundary between FM and FR-FM phases, $J=V, h \neq 0$, coincides with the magnetization for these phases: $M=n_{s}=1 / 2-m$. At the boundary corresponding to the spin flip field, $|h|=-2 J$, the magnetization is written as:
\begin{equation}
	M = \frac{1}{5} \left( 2 + 4m + \sqrt{\frac{1}{4}-9m+m^2} \right) . 
	\label{eq:ms-spin-flip}
\end{equation}
At the boundary corresponding to the frustration field, $|h|=-V-J$, we obtain
\begin{equation}
	M = \alpha \, \left(\frac{1}{2} - m\right) , 
	\label{eq:ms-frustr-field}
\end{equation}
where $\alpha$ is calculated from equation (42).

Figure 3 shows concentration dependences of magnetization for the phase states and states at the phase boundaries at $h \neq 0$. At the boundaries, the magnetization demonstrates non-linear dependence on $m$ and has an intermediate value compared with the magnetization of the adjacent phases. Magnetization (43) is shown in Figure 3,a. Its value for the pure spin chain is equal to $M_{0}=\frac{1}{\sqrt{5}} \approx 0.447$ and reaches its maximum $M_{\max }=4-2 \sqrt{3} \approx 0.536$ at $m=9 / 2-8 / \sqrt{3} \approx-0.119$.

\begin{figure}
\includegraphics[width=0.5\textwidth]{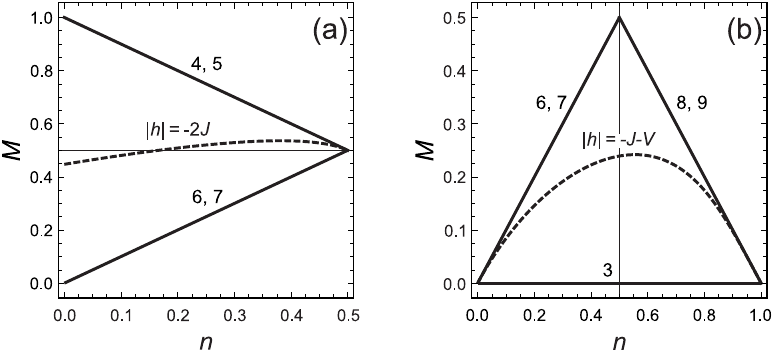}
\caption{Concentration dependences of magnetization. Numbers near the solid lines correspond to the states in Table I, equations near the dashed curves correspond to the ground state phase boundaries.}
\end{figure}

At $m=0, \mathrm{FR}-\mathrm{FM}$ and $\mathrm{FR}-\mathrm{FM}$ phases are transformed into FR-PM phase, so all three dependences merge into a single one, $M=n_{s}$. Magnetization (44) at the boundary between AFM and frustrated phases is shown in Figure 3, $b$. Note that this curve is not symmetric with respect to line $m=0$ and has its maximum $M_{\max } \approx 0.242$ at $m \approx 0.055$.

\section{Conclusions}
In the paper, we present the calculation and analysis of ground state phase diagrams of the dilute Ising chain in the magnetic field at the fixed impurity concentration; find correlation functions and properties of local distributions for frustrated phases, derive the general expression for the frustrated state entropy based on the Markov property of the system, and offer the calculation method for state entropy at the phase boundaries. These methods allow extension to other one-dimensional models with the Ising type interactions. 

It has been shown that the system ground state in the magnetic field remains frustrated, if the exchange interaction constant values satisfy the inequality $-|h|-V<J<V$. When the magnetic field increases in the AFM phase to the values higher than the frustration field $|h|=-J-V$, charge ordering occurs in the system. The maximum of this effect is observed when the spin chain is filled with impurities by half. Explicit concentration dependences of magnetization have been found for the phase boundary states that exhibit nonlinear behavior, while magnetization for the adjacent phases is linear by the impurity concentration. The obtained expressions for the residual entropy reproduce the known analytical results in the zero magnetic field which verifies correctness of the offered methods. It is shown that the residual entropy at the phase boundaries is always higher than the adjacent phase entropy, which proves that there are no pseudotransitions in the dilute Ising chain. For pseudotransition to occur, such system modification is required when the frustrated phase does not form mixed states with the low-entropy phase at the phase boundary.

\section*{Funding}
This study was supported by the Ministry of Science and Higher Education of the Russian Federation, project FEUZ2023-0017.

\bigskip
\section*{References}
{
\setlength\parindent{0pt}
\setlength{\parskip}{6pt}

[1] E.S. Tsuvarev, F.A. Kassan-Ogly. ZhETF 160, 2, 232 (2021). (in Russian).

[2] O. Rojas, J. Strečka, S.M. de Souza. Solid State Commun. 246, 68 (2016).

[3] I.M. Carvalho, J. Torrico, S.M. de Souza, M. Rojas, O. Rojas. JMMM 465, 323 (2018).

[4] I.M. Carvalho, J. Torrico, S.M. de Souza, O. Rojas, O. Derzhko. Ann. Phys. 402, 45 (2019).

[5] O. Rojas, S.M. de Souza, J. Torrico, L.M. Verissimo, M.S.S. Pereira, M.L. Lyra. Phys. Rev. E 103, 4, 042123 (2021).

[6] L. Gálisová, J. Strečka. Phys. Rev. E \textbf{91}, \textit{2}, 022134 (2015).

[7] S.M. de Souza, O. Rojas. Solid State Commun. \textbf{269}, 131 (2018).

[8] J. Strečka, R.C. Alécio, M.L. Lyra, O. Rojas. JMMM \textbf{409}, 124 (2016).

[9] O. Rojas, J. Strečka, M.L. Lyra, S.M. de Souza. Phys. Rev. E \textbf{99}, \textit{4}, 042117 (2019).

[10] O. Rojas. Acta Phys. Pol. A \textbf{137}, \textit{5}, 933 (2020).

[11] O. Rojas. Braz J. Phys. \textbf{50}, \textit{6}, 675 (2020).

[12] Y. Panov, O. Rojas. Phys. Rev. E \textbf{103}, \textit{6}, 062107 (2021).

[13] F. Rys, A. Hintermann. Helv. Phys. Acta \textbf{42}, \textit{4}, 608 (1969).

[14] M.P. Kawatra, L.J. Kijewski. Phys. Rev. \textbf{183}, \textit{1}, 291 (1969).

[15] F. Matsubara, K. Yoshimura, S. Katsura. Can. J. Phys. \textbf{51}, \textit{10}, 1053 (1973).

[16] Y. Termonia, J. Deltour. J. Phys. C \textbf{7}, \textit{24}, 4441 (1974).

[17] B.Y. Balagurov, V.G. Vaks, R.O. Zaitsev. Sov. Phys Solid State \textbf{16}, \textit{8}, 1498 (1975).

[18] A.V. Shadrin, Yu.D. Panov. JMMM \textbf{546}, 168804 (2022).

[19] Yu.D. Panov. JMMM \textbf{514}, 167224 (2020).
}

\end{document}